\documentclass{PoS}

\newcommand{\csh}{{\cos\theta^{*}}}
\newcommand{\sineff}{\sin^2\theta^{lept}_{eff}}
\title{Electroweak precision measurements with the CMS detector}
\ShortTitle{Electroweak precision measurements with the CMS detector}
\author{\speaker{A. Bodek}\thanks{CMS Collaboration, JINST 3 S08004 (2008).}
\\
        University of Rochester (on behalf of the  CMS collaboration)
\\         E-mail: \email{bodek@pas.rochester.edu}}

\abstract{
We report on a precision measurement of the effective weak mixing angle using the forward-backward asymmetry of Drell-Yan  ($ee$ and $\mu\mu$) events in pp collisions at $\sqrt{s}=8~\mathrm{TeV}$ at CMS. The data sample corresponds to an integrated luminosity of $18.8~\mathrm{fb}^{-1}$ and $19.6~\mathrm{fb}^{-1}$ for muon and electron channels, respectively. The sample consists of 8.2 million dimuon and 4.9 million dielectron events. With new analysis techniques and large samples the statistical and systematic uncertainties are reduced by a factor of two compared to previous measurements at the LHC. The extracted value of the effective weak mixing angle from the combined $ee$ and $\mu\mu$ data samples is 
$  sin^2\theta^{lept}_{eff}=0.23101\pm 0.00036(stat)\pm  0.00018(syst)\pm   0.00016(theory)\pm 0.00030(pdf)$   or $   sin^2\theta^{lept}_{eff}=0.23101\pm0.00052$.
 }

\FullConference{The European Physical Society Conference on High Energy Physics\\
		5-12 July, 2017\\
		Venice}

\begin{document}
\section{Introduction}

We report on a measurement\cite{SMP-16-007} of the effective weak mixing angle using the forward-backward asymmetry, $A_{FB}$, in Drell-Yan ($ee$ and $\mu\mu$) events  in pp collisions at $\sqrt{s}=8~\mathrm{TeV}$ at CMS. In leading order dilepton pairs are produced through the annihilation of a quark and antiquark to dileptons via the exchange of a  $Z$ boson or a virtual photon
% $\Pq\bar{\Pq}\rightarrow Z/\gamma^*\rightarrow \ell^+\ell^-$.%
 The definition of  $A_{FB}$ is based on the angle $\theta^*$ of the lepton ($\ell^-$) in the Collins-Soper  frame in the center of mass of the  dilepton system:
\begin{equation}
    A_{FB}=\frac{\sigma_{F}-\sigma_{B}}{\sigma_{F}+\sigma_{B}},
\end{equation}
where $\sigma_{F}$ and $\sigma_{B}$ are the cross sections in the forward ($\csh>0$) and backward ($\csh<0$) hemispheres, respectively. In this frame the $\theta^*$ is the angle of the $\ell^-$ direction with respect to the axis that bisects the angle between the direction of the quark and opposite direction of the anti-quark. In $pp$ collisions the direction of the quark is assumed to be in the boost direction of the dilepton pair.  In terms of  laboratory-frame energies and momenta  $\csh$  is equal to 
\begin{equation}
    \csh={\frac{2(p_1^+p_2^- - p_1^-p_2^+)}{\sqrt{M^2(M^2+P_{T}^2)} }}  \times\frac{P_z}{|P_z|},
\end{equation}
where $M$, $P_{T}$, and $P_{z}$ are the mass, transverse momentum, and longitudinal momentum, respectively, of the dilepton system, and $p_1(p_2)$ are defined in terms of the energy, $e_1 (e_2)$, and longitudinal momentum, $p_{z,1}(p_{z,2})$, of the negatively (positively) charged lepton as $p_{i}^\pm=(e_i\pm p_{{z},i})/\sqrt{2}$%~\cite{Collins}. 
A non-zero $A_{FB}$ in dilepton events originates from the vector and axial-vector couplings of electroweak bosons to fermions.

 The most precise  previous measurements of $\sineff$  are reported by  LEP and SLD experiments
%~\cite{ALEPH:2005ab}
 However, the two most precise measurements differ  by more than 3 standard deviations. Measurements of $\sineff$  are also reported by LHC\cite{ATLAS,LHCb}. and Tevatron experiments.
 %~\cite{Chatrchyan:2011ya,Aad:2015uau,Aaij:2015lka,Aaltonen:2014loa,Aaltonen:2016nuy,Abazov:2014jti}.
In this analysis we measure the leptonic effective weak mixing angle  ($\sineff$) by fitting the mass and rapidity dependence of the observed $A_{FB}$ in dilepton events. Statistical and systematic errors are reduced by using three new analysis techniques: (A) Angular event weighting\cite{eventw}, (B) Precise muon and electron energy calibration]\cite{momentumw} and
(3) Constraining PDF errors using the $A_{FB}$ dilepton samples (Bayesian $\chi^2$ reweighting of PDF replicas)\cite{eventw,momentumw,PDFw} .

\section{ Angular event weighted $A_{FB}$}

In the Collins-Soper frame the angular distribution of dilepton events has a  (1+$\cos^2\theta^*$) term that originates from the spin 1 of the exchanged boson, a $\csh$ term  from vector-axial interference and a $(1-3\cos^2\theta^*)$ term from the transverse momentum of the interacting partons
%~\cite{angular}.  
The angular coefficients  $A_0$ and $A_4$ are functions of dilepton mass ($M$,), transverse momentum  ($P_{T}$) and rapidity($y$)
of the dilepton pair. 
% In each dilepton mass and rapidity bin, the $\csh$ in full phase-space is distributed as
 %~\cite{angular}:
% add reference {angular} E. Mirkes and J. Ohnemus, Phys. Rev. D 50, 5692 (1994); 51, 4891 (1995)
\begin{equation}
\frac{1}{\sigma}\frac{{d}\sigma}{{d}\csh} = \frac{3}{8}\Big(1+\cos^2\theta^*+\frac{A_0 (M,P_T,y)}{2}(1-3\cos^2\theta^*) + A_4 (M,P_T,y)\csh\Big).
\end{equation}
In this analysis, the $A_{FB}$ values in each dilepton rapidity and mass bin are calculated using the ``angular event-weighting'' method, described in detail in Ref.~\cite{eventw}.
The technique is equivalent to measuring  $A_4$ in bins of $|\cos^2\theta^*|$, and extracting $A_{FB}$ from the average  $A_4$ for each dilepton mass bin. 

The ``angular event-weighted''  $A_{FB}$ is the same as the full phase-space $A_{FB}$, while the simple  fiducial restricted $A_{FB}$ ($A_{FB}^{restricted}$) values are smaller because of the limited acceptance at large $\csh$. Because of this feature, the event-weighted $A_{FB}$ is less sensitive to the exact modeling of the acceptance than $A_{FB}^{restricted}$. Additionally, because the event-weighted $A_{FB}$ exploits full shape of $\csh$ distribution as opposed to the sign only in the  case of  $A_{FB}^{restricted}$, it also results in  a smaller statistical uncertainty in $\sineff$.

\begin{figure}[!htbp]
\centering
      \includegraphics[width=10cm, height=7cm]{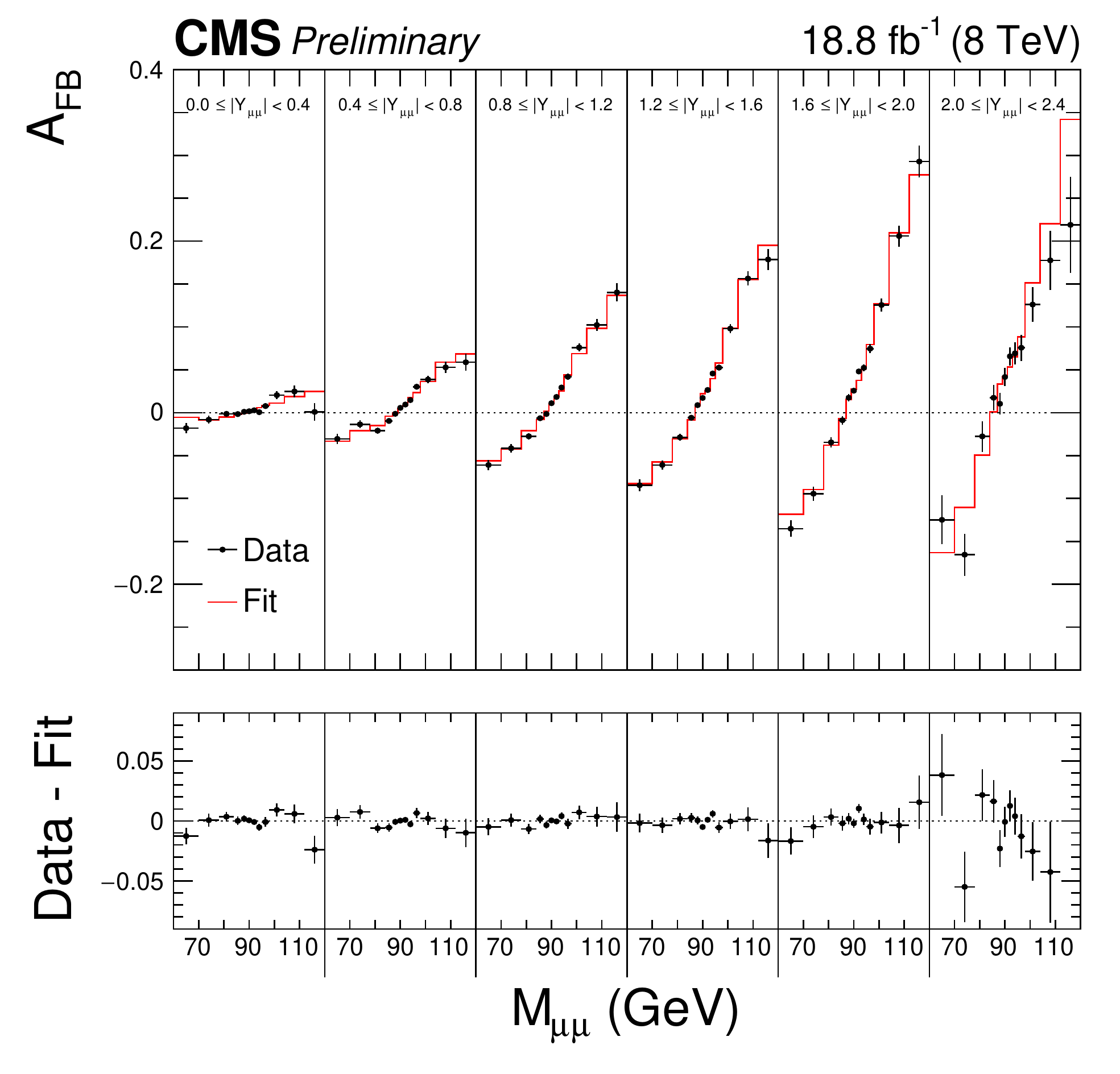}
        \includegraphics[width=10cm, height=7cm]{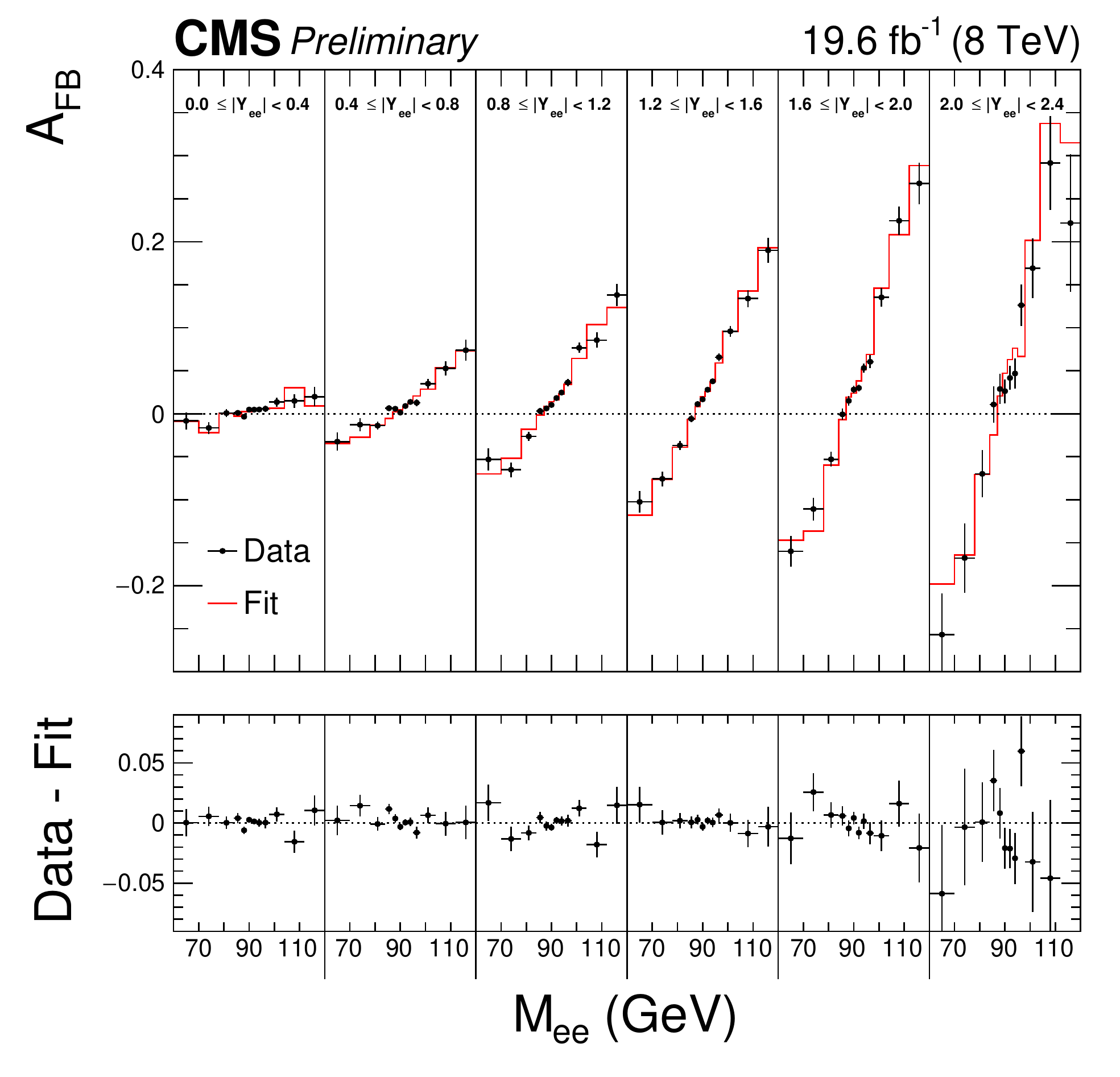}
    \caption{
	Comparison between the ``angular event-weighted'' $A_{FB}$ in data and the best-fit theory prediction for  $A_{FB}$ as a function of dilepton mass for  the dimuon (top) and dielectron (bottom) channels.  The best-fit theory prediction value for  b $A_{FB}$  each bin is obtained by linear interpolation between the two neighboring best fit templates in $\sineff$. 
	The templates are based on the central PDF of the NLO NNPDF3.0 set. \label{figure:fit}
    }
\end{figure}

\section{$\sineff$ extraction}

We extract $\sineff$  by minimizing the $\chi^2$ value between the data and template $A_{FB}$ distributions in 72 dilepton mass and rapidity bins. 
The default signal templates are generated with the $\textsc{powheg}$ event generator using the NNPDF3.0 set. $\textsc{powheg}$ is interfaced with $\textsc{pythia8}$
%~\cite{PYTHIA8} 
with CUETP8M1
%~\cite{tune}
 underlying event tune for parton showering and hadronization, including electromagnetic FSR. The template variations for different values of $\sineff$, renormalization and factorization scales, and PDFs are modeled using the  $\textsc{powheg}$ MC generator that provides matrix-element based event-by-event weights for each variation. To propagate these variations to the full-simulation-based templates, each event of the full-simulation sample is weighted by the ratio of $\csh$ distributions obtained with modified and default $\sineff$ configuration in each dilepton mass and rapidity bin. 

A comparison between the data and best-fit $\sineff$  template distributions is shown in Fig.~\ref{figure:fit}. 
Table~\ref{table:staterrors} summarizes the statistical uncertainty in the extracted $\sineff$ in the muon and electron channels and their combination. 
\begin{table*}[htbp]
\centering
%\topcaption{
\caption{
    Summary of the statistical uncertainties in the measurement of $\sineff$. 
    %The MC statistical uncertainties are after smoothing.
    The statistical uncertainties in the lepton selection efficiency and calibration coefficients in data are included.
    %The statistical uncertainties of the electron selection efficiencies and energy calibration coefficients, having no charge dependence, 
    %are much smaller and are evaluated separately. 
    \label{table:staterrors}
}
\begin{tabular}{ | l | r | r | }
\hline
channel    &  statistical uncertainty\\ 
\hline
muon		       &  0.00044  \\
electron	       &  0.00060  \\ \hline
combined	       &  0.00036  \\
\hline
\end{tabular}
\end{table*}
\begin{table*}[!htbp]
%\begin{table*}[]
\centering
%\topcaption{
\caption{
     \label{table2}
    Summary of experimental systematic uncertainties (A)  and theory modeling uncertainties (B) in the measurement of $\sineff$ in  the dimuon (left) and dielectron (right) channels. 
    For details see ref. \cite{SMP-16-007}.
}
\begin{tabular}{ l | c | c }
\hline
Source & muons & electrons \\
\hline
MC statistics		     & 0.00015 &	0.00033 \\
Lepton momentum calibration  & 0.00008 &	0.00019 \\
Lepton selection efficiency  & 0.00005 &	0.00004 \\ 
Background subtraction	     & 0.00003 &	0.00005 \\
Pileup modeling		     & 0.00003 &	0.00002 \\
\hline
Total	 experimental  systematic uncertainties		     & 0.00018 &	0.00039 \\
\hline\hline
Model variation &  Muons &  Electrons \\ 
\hline
Dilepton $P_T$ modeling				& 0.00003 &  0.00003 \\ 
QCD $\mu_{R/F}$ scale					& 0.00011 &  0.00013 \\ 
$\textsc{powheg}$  MiNLO Z+j vs NLO Z model				& 0.00009 &  0.00009 \\
FSR model ($\textsc{photos}$ vs $\textsc{pythia}$)				& 0.00003 &  0.00005 \\ 
UE tune							& 0.00003 &  0.00004 \\ 
Electroweak ($\sin^2\theta^{{lept}}_{{eff}} - \sin^2\theta^{{u, d}}_{{eff}}$ ) & 0.00001 &  0.00001 \\ 
\hline 
Total theory modeling uncertainties						& 0.00015 &  0.00017 \\ 
\hline
\end{tabular}
\end{table*}
%\section{Experimental systematic errors and theory uncertainties}

The systematic and theory modeling uncertainties are summarized in Table ~\ref{table2}. A detailed discussion of the  experimental systematic and theory modeling uncertainties are given in Ref. \cite{SMP-16-007}.  
\section{Constraining PDFs with $A_{FB}$ data and Bayesian $\chi^2$ PDF reweighting}
The observed  $A_{FB}$ values depend on the size of the dilution effect, as well as on the relative contributions from u and d valence quarks to the total dilepton production cross section. Therefore, the PDF uncertainties translate into sizable variations in the observed  $A_{FB}$ values.
However, changes in PDFs affect the $A_{FB}$($M_{\ell\ell}$, $Y_{\ell\ell}$) distribution in a different way from changes in $\sineff$. 

Changes in PDFs result in changes in $A_{FB}$  in regions where the absolute values of $A_{FB}$ is large, i.e. at high and low dilepton masses. In contrast,
 the effect of changes in $\sineff$ are largest near the Z-peak and are significantly smaller at high and low masses. Because of this behavior.
 %which is illustrated in Fig.~\ref{figure:pdftheory}, 
 we can apply the Bayesian $\chi^2$ reweighting method to 100 NNPDF3.0 PDF replicas to  constrain the PDFs~\cite {PDFw, Giele, Sato} and  reduce the PDF errors in the extracted value of $\sineff$.

The extracted $\sineff$ in the electron and muon decay channels and their combination with and without constraining the PDF uncertainties are shown in Table~\ref{table:combination}. 
 After Bayesian $\chi^2$ reweighting, the PDF uncertainties are  reduced by about a factor of 2. It should be noted that the Bayesian $\chi^2$ reweighting technique works well if the PDF replicas span the optimal value on both sides.  Additionally, the effective number of replicas after $\chi^2$ reweighting, $n_{{eff}}=N^2/\sum_{i=1}^{N}w_i^2$, should also be large enough to give reasonable estimate of the average and the standard deviation. The number of effective replicas after the $\chi^2$ reweighting is $n_{{eff}}=41$. As a cross check we also perform the analysis with the corresponding 1000-replica NNPDF set in the dimuon channel and find good agreement between the two results
\begin{table}[!htbp]
\centering
%\topcaption{
\caption{
    Central value and PDF uncertainty of the measured $\sineff$ in the muon and electron channels and their combination 
    with and without constraining PDFs using Bayesian $\chi^2$ reweighting.
    %The first uncertainty includes the combined statistical and experimental systematic uncertainties. 
    %The second  uncertainty is the NNPDF3.0 PDF uncertainty.  
}
\label{table:combination}
\begin{tabular}{ l | c | c }
\hline
Channel		   &  without constraining PDFs	    & with constraining PDFs \\ \hline 
Muon               &  $0.23125\pm0.00054$ & $0.23125\pm0.00032$  \\
Electron           &  $0.23054\pm0.00064$ & $0.23056\pm0.00045$  \\
Combined           &  $0.23102\pm0.00057$ & $0.23101\pm0.00030$  \\
\hline
\end{tabular}
\end{table}

\section{Summary}

We report on the extraction\cite{SMP-16-007} of  $\sineff$ from the measurements of the mass and rapidity dependence of $A_{FB}$ in Drell-Yan $ee$ and $\mu\mu$ 
events in pp collisions at $\sqrt{s}=8~\mathrm{TeV}$ at CMS.  
With larger samples and new analysis techniques (including precise lepton momentum calibration, angular event weighting, and additional PDF constraints from  Bayesian $\chi^2$ reweighting),  the statistical and systematic uncertainties are reduced by a factor of two compared to previous measurements at the LHC\cite{ATLAS,LHCb}. 
The combined result from the dielectron and dimuon channels is:

\begin{eqnarray}
	\sineff&=&0.23101\pm
            0.00036({stat})\pm
	    0.00018({syst})\pm
	    0.00016({theory})\pm
	    0.00030({pdf}) \nonumber \\
	\sineff&=&0.23101\pm0.00052. \nonumber
\end{eqnarray}

\begin{figure}[ht]
    \includegraphics[width=10cm, height=8cm]{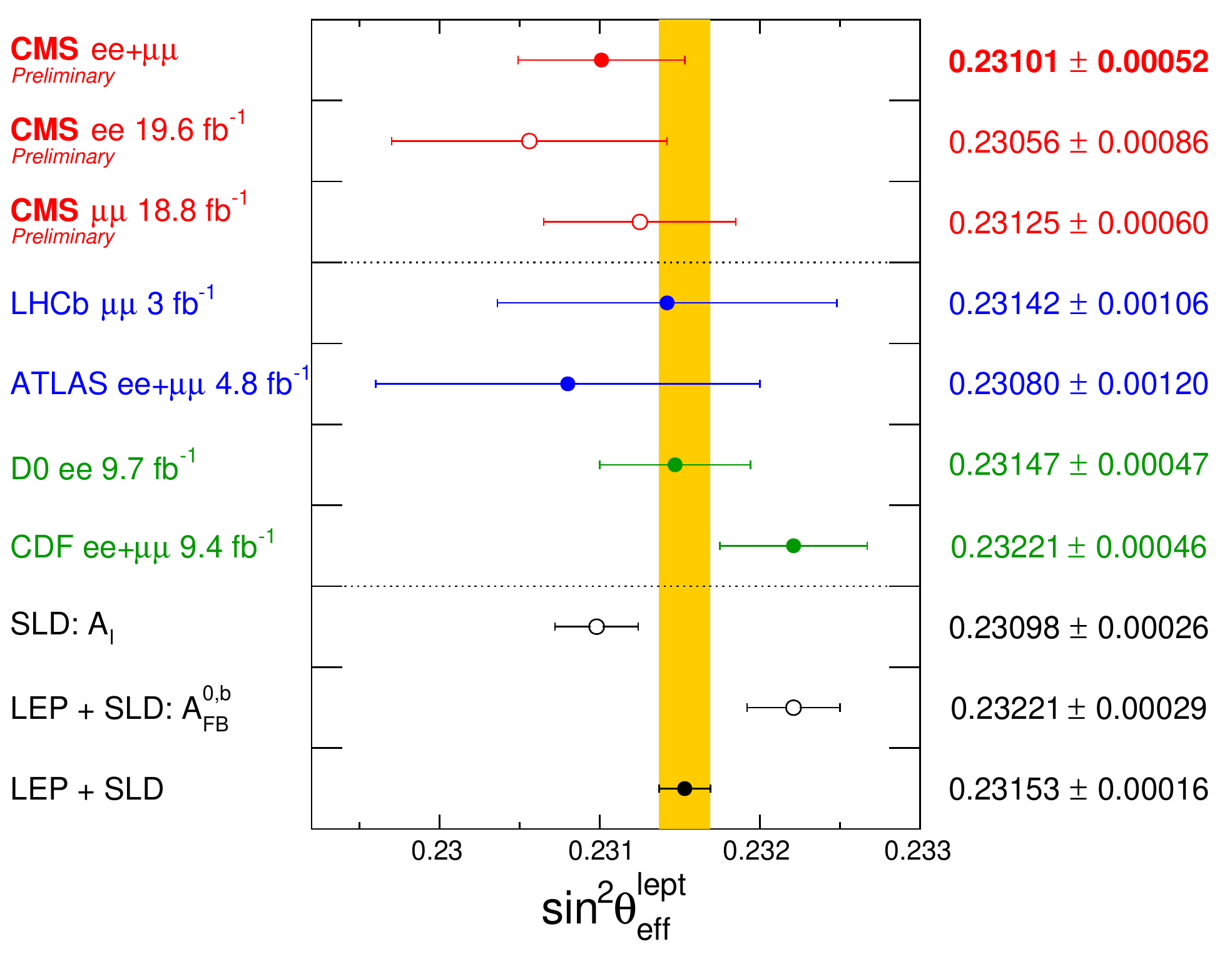}
    \caption{
	Comparisons of the measured CMS $\sineff$ in the muon and electron channels and their combination with
	previous LEP/SLC, Tevatron and LHC\cite{ATLAS,LHCb}. measurements.  
	The shaded band corresponds to the combination of the LEP and SLC measurements. 
	\label{figure:result}
    }
\end{figure}

Comparisons of the extracted $\sineff$ with previous results from  LEP/SLC, Tevatron and LHC\cite{ATLAS,LHCb}. are shown in Figure~\ref{figure:result}. 
The results are consistent with the most precise LEP and SLD measurements.

\end{document}